# Rescribe: Authoring and Automatically Editing Audio Descriptions

**Amy Pavel**
Apple Inc.
apavel@apple.com

**Gabriel Reyes**
Apple Inc.
gareyes@apple.com

**Jeffrey P. Bigham**
Apple Inc.
jbigham@apple.com

## ABSTRACT
Audio descriptions make videos accessible to those who cannot see them by describing visual content in audio. Producing audio descriptions is challenging due to the synchronous nature of the audio description that must fit into gaps of other video content. An experienced audio description author will produce content that fits narration necessary to understand, enjoy, or experience the video content into the time available. This can be especially tricky for novices to do well. In this paper, we introduce a tool, Rescribe, that helps authors create and refine their audio descriptions. Using Rescribe, authors first create a draft of all the content they would like to include in the audio description. Rescribe then uses a dynamic programming approach to optimize between the length of the audio description, available automatic shortening approaches, and source track lengthening approaches. Authors can iteratively visualize and refine the audio descriptions produced by Rescribe, working in concert with the tool. We evaluate the effectiveness of Rescribe through interviews with blind and visually impaired audio description users who give feedback on Rescribe results. In addition, we invite novice users to create audio descriptions with Rescribe and another tool, finding that users produce audio descriptions with fewer placement errors using Rescribe.

## INTRODUCTION
Audio descriptions (AD) make videos accessible by describing important visual content in the audio for those who cannot see it. Well-crafted audio descriptions that appear alongside movies, TV, and streaming services provide minimal distraction from source content through mixing and placement that avoids overlapping speech and important sounds [13]. Audio descriptions that play synchronously alongside the video are known as *inline audio descriptions*. To create such descriptions, audio description authors or teams identify regions where descriptions may fit, identify visual content to describe, script the audio descriptions, record the voiceover, and edit the voiceover into the source track or accompanying narration track. Each task requires effort and specialized skill such that studios may employ audio description teams of script writers, audio engineers, voiceover actors, and producers to craft descriptions.

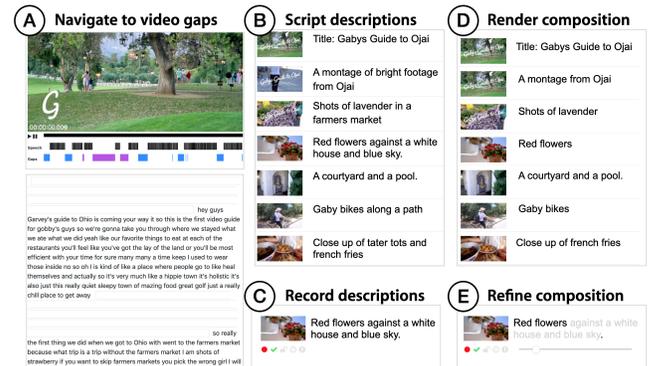

**Figure 1.** Rescribe enables audio description authors to efficiently create audio descriptions. Using Rescribe, authors navigate to speech gaps in the timeline or transcript, and reference the transcript for description context (A). Authors then script and record their descriptions (B,C). Rescribe then renders a composition of their descriptions with the source track, automatically editing the descriptions to fit within the gaps (D). Finally, authors can edit and refine their descriptions (E).

A key challenge in creating inline audio descriptions is fitting all of the content important to comprehension within the limited space available in the video. In order to maximize use of the time provided, authors use trial-and-error to iterate through options for the placement and content of audio descriptions to shorten what they want to say in the time they have to say it. While this iterative process can be tedious and time-consuming for professionals and novices alike, new audio describers can exacerbate this problem through over-description (e.g., adding information such as camera angles or coloring that is less important to comprehension). Non-professional describers can turn to *extended descriptions* [25] that pause the audio track to allow for more description with the trade off of disrupted source audio and increased playback time.

To allow describers to efficiently produce audio descriptions that maintain the length and audio quality of the existing video, we present Rescribe. Using Rescribe, authors can navigate to gaps in the source video dialog (Figure 1A), then script and record (Figure 1B,C) all of the content they would like to include in their descriptions. Rescribe then uses a dynamic programming approach to optimize between the placement of the descriptions and available automatic shortening approaches (e.g., word selection and sentence shortening). Authors can iteratively visualize and refine the audio descriptions produced by Rescribe, working in concert with the tool. Rescribe also lets authors create a new form of audio descriptions, *extended-inline descriptions*, that lengthen the source audio content by looping the music to preserve the video flow (Figure 2). To create extended-inline descriptions, Rescribe additionally considers source track lengthening approaches (e.g., extending silence, or looping music) alongside description placement, and shortening options within the optimization. Using one set of draft descriptions, authors can export all three types of descriptions: extended descriptions, inline descriptions, and extended-inline descriptions.

We share a set of Rescribe-generated results with 7 blind and visually impaired audio description users, and find that all users prefer extended-inline and inline descriptions to extended descriptions, and expressed excitement about using extended-inline descriptions in the future. We also compare using Rescribe with a traditional audio description interface with 8 users and find that Rescribe lets users efficiently create inline audio descriptions with fewer placement errors (reducing an average of 45% of descriptions overlapping with dialog content or other descriptions to 0%) than the traditional interface. Rescribe also lets users achieve a higher number of descriptions within the available gaps compared to a traditional timeline-based editor (Final Cut Pro). We conclude with initial impressions from professional audio describers who suggest uses for Rescribe within their workflow.

In summary, this work contributes:

- Rescribe, a system for authoring and automatically editing audio descriptions
- A new form of audio descriptions, *extended-inline* descriptions, made possible by Rescribe
- Study results comparing Rescribe to a timeline-based editor for creating audio descriptions
- An interview study with blind and visually impaired users comparing extended, inline, and extended-inline audio descriptions

## BACKGROUND

Our work builds upon prior work in creating audio descriptions, writing support, and media editing based on constraint solving.

### Describing videos

Audio descriptions are challenging to produce, due to both the careful timing of descriptions in between existing audio content, and the multitude of production skills required for the task (e.g., scripting, audio engineering, voice acting). While professionals often use a mix of captioning tools (e.g., InkScribe) and audio editing software (e.g., Logic), prior work has proposed authoring tools to facilitate the creation of audio descriptions. 3PlayMedia [5], YouDescribe [25], LiveDescribe [9], and

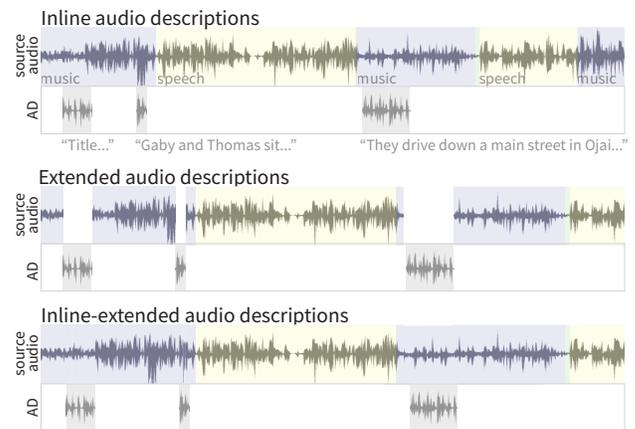

Figure 2. Inline audio descriptions synchronously describe the content of the video while the source audio plays back. Well-crafted inline descriptions avoid important audio segments (e.g., dialog). Extended audio descriptions pause the playback of the source audio such that the extended descriptions always disrupt the source audio and never overlap important audio. Our extended-inline descriptions enable authors to extend the source material to accommodate additional description without disrupting playback.

Gagnon et al. [18] offer timeline-based editors for creating audio descriptions.

3PlayMedia's post-production audio description tool [5] and Gagnon et al.'s audio description tool [18], both let authors provide text descriptions on videos, synthesizing text-to-speech for playback of the descriptions. Gagnon et al.'s tool also provides authors with timeline-based visualisations tailored to the production of cinematic audio descriptions including recognition of scenes, characters, and important locations [18]. While prior work suggests that people prefer human-narrated audio descriptions to speech-to-text audio descriptions when available [15, 28], the aforementioned systems do not support narrated audio descriptions. Two tools that do allow for spoken audio descriptions are LiveDescribe [9] and YouDescribe [25]. YouDescribe and LiveDescribe are both timeline-based voiceover recording and editing tools that allow people to script and record audio descriptions. LiveDescribe visualizes predicted gaps along the timeline, but authors still must iteratively refine their descriptions to fit within the space provided. We build upon such work by assisting authors in the process of iteratively refining their recorded descriptions by optimizing the audio description and source track. Rescribe also surfaces the transcript to aid in estimating gap length, finding description vocabulary, and reflecting on the coverage of audio descriptions with the source video narration (e.g., ensuring minimal repetition of information that can be inferred).

### Writing support tools

History of work considers how to support humans with their writing tasks. Spell check represents an early success of writing support tools [43], and grammar correction remains an active area of research [62] and product development [4]. Low level suggestions such as word-level ("fecundity" to "fertility" [27]) and phrase-level ("I am hoping" to "I hope" [4]) meaning-preserving simplifications are also used to support writing. To shorten an author's draft descriptions to fit within the time provided, Rescribe draws from the active research areas of extractive summarization [37] (identifying important information in a document) and sentence compression (shortening sentences) [16], which are less commonly used in existing tools for writing support. While prior work reveals people can collaboratively perform paragraph shortening [7] and summarization [63, 58] tasks with success, automatic techniques still achieve varying results, especially with respect to common focus areas of drastic text reduction (e.g., headline generation, bullet point creation). Rescribe mitigates challenges with automatic summarization by conservatively instantiating concepts from extractive summarization (e.g., importance, efficiency) and sentence compression (e.g., using dependency trees) in the task-specific context of audio description.

Campos et al. and Lakritz et al. similarly consider scripting audio descriptions, but for cinematic content where movie scripts exist [10, 31]. While not intended for audio descriptions, scripts often contain scene description and Campos et al. identify what subset of script sentences are likely to make good audio descriptions (e.g., using frequently used words in audio descriptions, character names, common words in the script), greedily placing them in available gaps, then applying text to speech. Rescribe works in the common case for independently created videos where a script does not exist (or does not reflect the final edit) by taking authors' draft descriptions (intended for inclusion in audio descriptions), their placement timing, and timing of the description recording into account to make subtle changes that preserve author intent. Towards this goal, Rescribe uniquely uses sentence compression (e.g., removing adjectives, short clauses, and prepositional phrases) to make less noticeable scripting changes, and optionally considers audio track lengthening to preserve additional content.

**Text and optimization-based media editing**
Traditional editing tools like Premiere [44] typically use frame-by-frame navigation using video timelines. Several systems have used time-aligned transcripts to support audio / video editing through text manipulation operations [8, 11, 17, 24, 32, 41, 48, 50, 52, 55, 61]. Such tools provide, for example, clip segmentation via the transcript [41, 42], scripted narration support [50, 52], and transcript-based speech editing [8, 17, 49]. Prior work also uses such text-aligned media for selecting optimal footage or camera angles to go along with a script [32, 52, 55], for automatically generating an optimal score to accompany audio stories [47], or for controlling animated characters [54]. While such prior work explores automatic video and audio editing through manual text manipulation, such work does not yet explore automatic text manipulation. Rescribe builds upon this prior work through a the exploration of a new domain, creating a new space of design opportunities in editing text along with the source media.

## DESIGNING FOR AUDIO DESCRIPTIONS
The Rescribe design is based on the established literature of guidelines for audio descriptions created by and/or in collaboration with visually impaired authors [1, 6, 13, 22, 25] and through talking to and reading content written by audio describers about the task. Here, we summarize key guidelines (denoted by G#) for inline audio descriptions from 5 sources selected for comprehensiveness and diversity (American Council of the Blind's guidelines [1], DCMP's Description Key [13], ADLab's guide for describing film [6], HHS's guidance [22], and YouDescribe's tips for description [25]):

**Content.** Describe important visual content essential to comprehending the video (G1). Avoid describing content that can be inferred from the audio, but do describe unidentified sounds (G2). Start general then progress to details (G3).

**Placement.** Avoid overlapping audio descriptions with dialog in the video (G4). Do not "spoil" the video by describing surprising content before it appears (e.g., a jump scare) (G5).

**Style.** Match the tone and the vocabulary of the source video and audience (G6). Be clear, concise, and use active voice when writing (G7). Be objective and avoid editorializing (particularly important in American guidelines [39]) (G8). When transcribing text, transcribe verbatim (G9).

**Preparation.** Script descriptions (G10), and familiarize yourself with the video (G11). Voiceover artists should practice their descriptions (G12).

Extended audio descriptions, audio descriptions in which the source content pauses to play back a description, are rarely referenced in guidelines, but players such as YouDescribe [25] and 3PlayMedia [5] support extended descriptions. YouDescribe suggests authors "use extended description when necessary", and WCAG 2.0 also references that such descriptions are used "when there is not enough space to include necessary narration between the natural dialogue" [59]. We aim to instantiate guidelines in our system and reference them inline during the system description. We leave some guidelines (e.g., maintaining objectivity, not overlapping with non-dialog important audio content, helping users identify visual content vocabulary not stated in the video) for future work, and allow support for users to execute such guidelines.

We additionally spoke to 7 audio description (AD) professionals (6 AD creators on a team, and 1 independent AD creator / instructor) to identify existing workflows and challenges. The AD team consisted of audio engineers, script writers, and producers. People who scripted audio descriptions reported 2 primary challenges: selecting what to convey in limited time available, and identifying and applying the correct vocabulary for the source material (e.g., according to the content context, or according to a third party). As one online writer summarizes: *"In producing good descriptions it's all about 'time'"..."I have to get in and out without stepping on the narrator. But obviously the description has to be comprehensible."* [60]. The AD instructor stated that fitting important content in limited space was particularly challenging for novice volunteers, and she preferred for novices to avoid thinking about timing when starting to describe (e.g., by using extended descriptions). On the other hand, AD creators that worked closely with source video production teams highlighted the importance of not altering video playback due to workflow compatibility and preserving author intent.

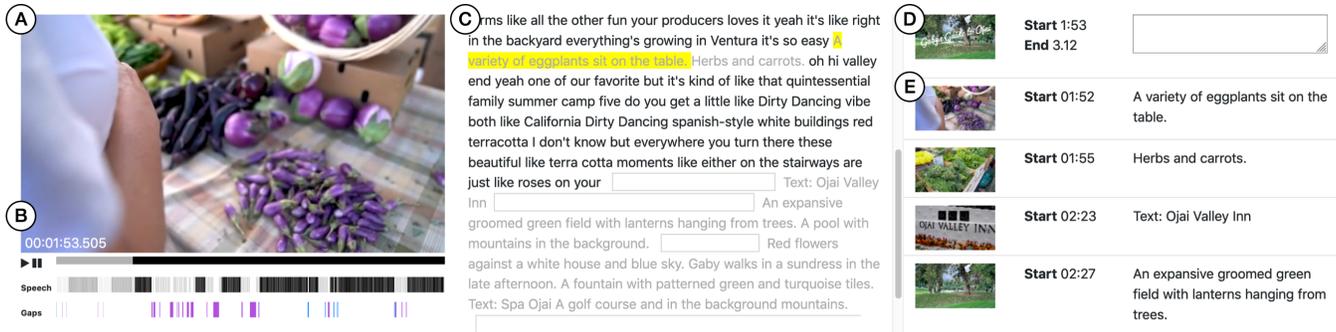

Figure 3. The interface features: 1) a video and timeline pane (left) with the source video speech in black and audio descriptions in grey (A,B); 2) a transcript pane (center) that visualizes the source video transcript (black), the audio descriptions (grey), and the gaps available (white) (C); and 3) the description pane (right) displaying the descriptions and thumbnails (D,E). As the video plays, the interface scrolls to the relevant point in the transcript (highlighted) and descriptions. Source video: *Gaby's Guide to Ojai* by What's Gaby Cooking [12].

## SYSTEM

To help video authors draft, record, and edit their audio descriptions Rescribe provides: 1) a video timeline for navigating the video, 2) a transcript for referencing the original terminology, and 3) a description pane for authoring, visualizing, and editing the audio descriptions.

The *timeline pane* lets the video description author play, pause, and seek within the source video. To guide authors in placement of their annotations, the timeline pane initially displays the content of the audio track (G4). Authors can avoid speech using the speech timeline, and locate good candidate placement regions using the gap timeline (gaps also visualized in LiveDescribe [9]). The two colors in the speech timeline indicate the source video speech (black), and the descriptions (grey). The two colors in the gap timeline indicate music gaps (blue) and extendable gaps (purple). The speech and gap timelines update as the author writes and edits their descriptions to show the remaining gap time. For each written description, the speech timeline updates to display the description placement and approximate description duration (based on an estimate of 0.3s per word), and the gap timeline updates to reduce the time remaining in the corresponding gap. When the author plays the video after writing descriptions, the system plays a preview of the descriptions using text-to-speech alongside the source audio.

The *transcript pane* displays a time-aligned transcript of the source and gaps in the form of spaces in the transcript. Clicking on a word or gap in the transcript navigates to the corresponding location in the video and descriptions. While writing audio descriptions, the transcript pane lets authors reference the terminology used in the original video (G6) to describe visuals that may not be clear from the video frame alone. For instance, a shot in a travel video shows a spread of indiscernible pub food, but the contents do not become clear until the traveller later mentions that she enjoyed the "wings", "cracklings", and "bibimbap". In addition, authors can use the transcript to see what has already been described in speech to avoid redundancy (G2). As the author writes or edits audio descriptions, the transcript pane displays their descriptions (in grey) alongside the original source transcript (in black), and reduces the size of the corresponding gap.

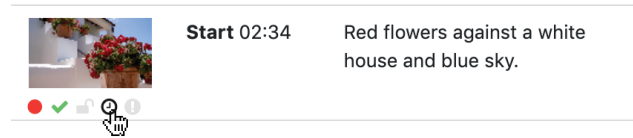

Figure 4. A user can select description rendering options by mousing over the description. From the left, the user can click the record button to record a description and the check mark will turn green once at least one description is recorded. The user can select to lock the description text (lock icon), the description time (clock icon), or the presence of description (exclamation point icon).

The *description pane* lets authors script and record their descriptions, and displays all of the descriptions for the video (G10). The descriptions are time aligned and navigable such that clicking on a description navigates to the corresponding location in the video and transcript. An author adds a description by typing their description into the current description tab. The system visualizes the start frame of the description in the thumbnail. By default, descriptions are previewed in speech-to-text, but the author can add a narrated description by clicking on the record button. To achieve a high-quality performance, authors can record multiple times and the system takes only the last recording created for a description (G12). Authors can also lock the description text to ensure an exact match (e.g., in the case of a text-on-screen transcription, G9) by either pressing a lock icon or by adding quotation marks to their description (Figure 4). Authors can also lock the description to a particular time (G5) by clicking on the clock icon (e.g., in cases where description should appear such as a translation of speech), or guarantee the presence of description in the final edit by clicking the exclamation point (e.g., for important moments such as introducing an interviewee). Finally, authors can edit the text or time of the description by double clicking, and delete descriptions using option-click.

After video description authors create a draft of the descriptions the author can select rendering options (e.g., inline, extended, or extended-inline descriptions) and click render to apply automatic adjustments. Once the description renders, the rendered audio track will appear under the main video track and the author can edit the new descriptions either by directly editing the description text, by selecting words for inclusion or by adjusting a slider to select a new description (Figure 5). When using word-based or slider-based editing, the original audio is edited such that the user does not need to re-record.

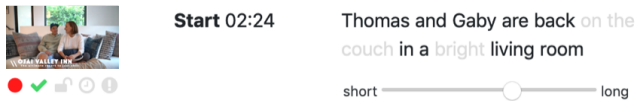

**Figure 5.** After Rescribe renders the new description, the user can access additional editing modalities through mouseover. While the original options remain, users can re-record by clicking on the record button. The included text is black, and the original text not included is grey. Users can slide to select an option automatically or click words in order to toggle their inclusion. Users can rewrite by double clicking and the recording will be deleted.

| | |
|---|---|
| **Original description** | |
| People walking along a beach with an overcast sky above and white sand below turquoise water . | |
| **Candidate descriptions ranked by coherence** | |
| 38.26 | People walking along a beach with an overcast sky and white sand . |
| 59.17 | People walking along a beach with white sand . |
| 67.73 | People walking along a beach . |
| 68.79 | People walking along a beach with an overcast sky above . |
| 1211.56 | People walking with an sky . |

**Figure 6.** Examples of candidate descriptions ranked by coherence where a lower score correspondes to a higher level of coherence.

## ALGORITHMIC METHODS

Rescribe relies on automatic methods to identify candidate description locations, generate revisions per description, align a selected revision to the audio track, and to generate the final audio track.

### Identifying description locations

To guide users and our algorithm in placing audio descriptions, we identify regions where audio descriptions may be added without disrupting the existing audio. Similar to prior work by Gagnon et al., we label non-speech regions [18]. To find regions that avoid distracting or important background sounds, we also identify regions of music and regions of ambient or background sound. First, we apply Doukhan et al.'s CNN-based music and speech segmentation to classify regions in audio as music, speech, and silence [14]. We also align words from the source video transcript (e.g., from existing captions or rev.ai) to the original audio, by first grouping together continuous captions following prior work [40], then using Gentle [38] alignment which finds an alignment between the phonemes in the words of the transcript and features in the audio. We label audio segments that are aligned to transcript words as speech, label the remaining audio as music or speech based on the classification. We label silence as extendable, and classify music as extendable if the segment is $\geq$ 30s and beat detection [34] detects beats at a rate $\geq$ 60bpm. We display the classifications in the interface and use them in the algorithm to inform placement of descriptions.

### Generating candidate descriptions

To simplify audio descriptions, we follow an approach similar to prior work in sentence simplification by first generating simplified candidates for each description based on the parse tree and later ranking the simplified candidates [57]. We constrain our generated candidate descriptions to ones that contain a subset of the words contained in the original description, such that we can later automatically generate the audio for a candidate description by concatenating existing audio for each word (i.e., without re-recording), as human narrated descriptions are preferred by users [28]. In order to generate subset candidate descriptions, we first parse the descriptions to find parts of speech and dependencies for each word in the description (using SpaCy's part of speech and dependency parser). We consider candidate tokens to drop as all adjectives, prepositional phrases, and compound phrases can typically be dropped without requiring further revision for grammatical correctness (e.g., conjugating a verb). We seek to maintain consistency of phrasing between the original description, the video, and the current description. We restrict to candidates that do not separate words in film-specific phrases ("cut to"), video-specific phrases (e.g., "big red dog"), phrases contained in quotes, or phrases specified by the user to be on-screen text (e.g., "Text on screen: Welcome back"). We determine film-specific references as any n-gram present in a set of pre-defined film terms (from an existing glossary of film terms [3]), followed by any number of prepositions (e.g., "zoom in" and "zoom in on" are film-specific phrases for the base word "zoom"). We determine video-specific phrases as any 2, 3, or 4-gram that (1) does not contain stop words, (2) occurs 3 or more times across the transcript and descriptions, and (3) includes at least one verb or noun. We determine phrases specified to be on-screen text as any phrase preceded in the description string by any of the sub-strings: "text", "title", or "credit". In the future, OCR could determine on-screen text.

### Scoring audio description compositions

We express an audio description composition as a series of descriptions $D_n = (d_1, ... d_n)$, one for each author-generated annotation, where $d_i$ is a tuple $(t, l)$ that specifies the description start time in the final composition, $t$, and the duration of the description, $l$. Our composition cost $E(D_n)$ seeks to quantify how well the description composition $D_n$ adheres to guidelines for the content and placement of audio descriptions. We quantify the cost of a given $d_i$ in terms of the quality of the text selected for a given description, $C(d_i)$, and the quality of the placement of the description in the audio description, $P(d_i)$. The overall cost of the audio description is:

$$E(D_n) = \sum_{i=0}^{n}(C(d_i) + P(d_i))$$

The candidate cost, $C(d_i)$, encodes the quality of the author-generated annotation and its alternatives in terms of the coherence, informativeness, and edit quality of the candidate.

**Coherence score.** Generating candidate descriptions using a task-tailored rule-based method brings the drawback of occasionally generating grammatically incorrect, or illogical descriptions. For instance, for the description "People walking along a beach with an overcast sky above and white sand below turquoise water" our system generates correct alternatives (e.g., "People walking along a beach with white sand", "People walking along a beach"), but also odd and grammatically incorrect alternatives (e.g.,"People walking with an overcast sky", "People walking with an sky"). To balance the sometimes erratic behaviour of our rule-based approach, we score the coherence of the candidate description sentences using GPT-2, a large pre-trained language model [45]. Specifically, we use the candidate description as input and directly use the output log loss of the model, which estimates the probability of the word sequence, as the coherence score.

**Informativeness score.** Within-sentence coherence does not consider the utility of the description with respect to the video and the surrounding descriptions. For instance, coherence alone ranks candidates for the annotation "Close up of bar food including bibimbap and tater tots" in the following order from least costly to most costly: "Close up", "Close up of bar food including tater tots", "Close up of bar food including bibimbap". The narration of the source video later discusses the bibimbap in detail rendering the reference to bibimbap important, despite a lower coherence score for the phrase containing "bibimbap". To encourage candidates with important terms, we first parse the candidate text, all author-generated descriptions, and the source video transcript using the method of Honnibal et al. (included in SpaCy [23]). We then score each noun and pronoun in the candidate text by dividing the number of times the word occurs in all author-generated descriptions and the source video text by the log probability of the word occurring in a general corpus (Spacy's Core Medium English corpus). Words important to the video (e.g., "bibimbap", "Ojai") receive higher scores than generally common words (e.g., "tree"), and we sum the scores over all nouns and divide by the number of words to get the informativeness score. We turn the informativeness score into a penalty by using the reciprocal. We consider author-generated descriptions in addition to the source video text, as character names or locations may appear only in descriptions. By considering nouns only, we avoid rewarding adjectives which appear more often in novice-generated descriptions. We also omit film language (any 1-gram contained in an existing film glossary [3]) from the list of important words, as they rarely provide important extra information to the listener. The importance score is similar to "the most frequent words in the script" score used by Campos et al. [10]. We additionally considered part of speech and film language to avoid common novice mistakes.

**Edit quality score.** We use word-level time alignment (from Gentle [38]) between the description text and the recorded audio description to edit the original descriptions to the shorter description candidates. Slight alignment errors and changing prosody of user speech can result in small editing artifacts in the final audio track. To minimize the potential for artifacts, we assign each candidate a penalty equal to the number of required cuts to the audio track. For instance, shortening "A long bench with blue birds" to "A bench with birds" or to "A long bench" requires 4 cuts or 1 cut, respectively (cuts displayed in red below). To the cut count penalty, we add a

| A|long|bench with|blue|birds |   | A long bench|with blue birds |

penalty of 20 (last word penalty) if the final word is removed from the description as the final word often contains downward intonation.

Overall, our candidate cost for any description candidate, $C(d_i)$, is the sum of the coherence, informativeness, and edit quality scores ($C_{coh}$, $C_{info}$, and $C_{edit}$) multiplied by their weights. We empirically set the weights $w_{coh}$, $w_{info}$, and $w_{edit}$ to 1, 500, and 10 respectively to balance the impact of each term. After weighting, all values have a range of approximately 300 for most candidates for a given description, with $w_{coh}C_{coh}(d_i)$ giving high scores (up to 30000) in the case of improbable (and often low-quality) candidates.

$$C(d_i) = w_{coh}C_{coh}(d_i) + w_{info}C_{info}(d_i) + w_{edit}C_{edit}(d_i)$$

To allow our algorithm to skip a description $d_i$ without placing any description candidate, we set $C(d_i)$ equal to 10000, a value selected to be high but not infinite, for all $d_i = (t, l)$ when $l$ is lower than the minimum length for all generated description candidates. For example, if the minimum generated description candidate for $d_i$ is 3 seconds long, $C(d_i)$ would be equal to 10000 for all $d_i$ only where $l < 3$.

Our placement score, $P(d_i)$, encodes the validity and quality of the placement of $d_i$ within the composition, including the overlap of speech in the original track, overlap of a previously placed audio description, and quality of non-speech placement.

**Overlap score.** We set a penalty for overlapping with speech in the source track, or a previously placed audio description ($d_0...d_{i-1}$), to infinity. We apply a penalty of 10 for placements within 0.3s, or approximately the length of a word, of an overlapping segment to encourage distinct descriptions. We apply the same penalty for descriptions that fall into regions without a silence or music label.

**Extension score.** We consider the extension score only when rendering extended-inline descriptions. For each extended description tuple $(t, l)$, we compute the extension score as the time in seconds that the gap would be extended to add $l$ at position $t$. The small penalty based on seconds allows the algorithm to extend the gaps where necessary, but otherwise maintain the original length. If the gap is not extendable music or silence, the score is set to infinity.

In total, our cost for any placement of a description $d_i$ with a placement of position $t$ in the track and the length of the description $l$ is simply the sum of the associated placement penalties computed for each audio window from $t$ to $t+l$:

$$P(d_i) = \sum_{j=t}^{t+l} P_{overlap}(j) + P_{extension}(d_i)$$

**Optimizing audio description compositions**
To find $D_n$ that minimizes the cost $E(D_n)$, we can express our audio description composition cost in terms of its recursive form $E(D_i) = E(D_{i-1}) + C(d_i) + P(d_i)$ allowing us to apply dynamic programming to achieve the globally optimum solution. For each user-generated annotation, $d_i$ ($0 \leq i \leq N$), we construct a table of possible solutions $(t, l)$ representing video placements, $t$, where $0 \leq t \leq M$ and $M$ is the video length, and proposed candidate text lengths, $0 \leq l \leq L$, where $L$ is the maximum candidate length over all candidates. We select the minimum of past and current costs at each step. At the final description $d_n$, we trace back from the $(t, l)$ with the minimum cost to find the solution that optimally minimizes the cost $D_n^*$. The runtime of our algorithm is then $O(NML)$ where $M$ is large. We constrain our search space to solutions where the placement of audio descriptions falls within 2 minutes and at least 1 shot (using FFMPEG's scene change detection [2] with the default threshold of 15) away from the original start time of the solution. As descriptions farther from their original time typically decrease in relevancy (except, perhaps when within the same shot), this constraint typically results in skipping descriptions that may otherwise be placed at a less relevant time.

**Rendering audio descriptions**

We use either text-to-speech or the author's recorded voice to serve as the voiceover audio for audio descriptions. To produce extended-inline descriptions, we first extend the relevant music segments of the audio track using the method proposed by Rubin et al. to retarget the length of a song by locating key transition points to generate loops [47]. When looping segments of ambient noise or silence, we avoid a looping approach as small artifacts become salient, sounding like beats upon regular repetition. Instead, until we have extended the segment to the desired length (no more than 2x its original length in our case), we randomly sample extraction points and 0.2-1.0s segments and then append each segment with a 0.0-0.2s offset from the end of the existing segment.

**EVALUATIONS: COMPARING AUDIO DESCRIPTIONS**

Informed by audio description guidelines, Rescribe automatically converts extended descriptions written by users into inline descriptions that fit within important audio. However, while extended audio descriptions do exist [9, 25, 28, 59], few guidelines exist to inform their use [25, 59], especially compared to the number of guidelines for the more typical inline audio descriptions. Also, no existing work directly compares extended descriptions with similar inline descriptions. In addition, Rescribe enables novel extended-inline descriptions for which guidelines of use do not exist. We aimed to assess: *Would people who use audio descriptions prefer to use inline or extended-inline descriptions over extended descriptions?*

**Materials:** Using Rescribe, an author of this paper drafted audio descriptions for a 9 minute and 36 second video ("Gaby's Guide to Ojai", a travel guide to Ojai, California [12]) and a third-party recorded the voiceover. We instrumented Rescribe to record interactions and identified that scripting took 20.6 minutes and recording took 8.0 minutes to produce the draft descriptions. From these draft descriptions, we produced 3 versions of audio descriptions from Rescribe: extended, inline and extended-inline. The extended descriptions matched the length of the drafted descriptions, the inline descriptions reduced description length by 52% of the original time while preserving 94% of description segments (e.g., by removing paraphrases, adjectives, and phrases), and the extended-inline descriptions reduced description length by 76% compared to extended, keeping 97% of description segments.

**Procedure:** We conducted remote semi-structured interviews with 7 blind and visually impaired participants to compare ver-sions of audio descriptions created with Rescribe. During each 30 minute interview, we first asked participants a few demographic questions (e.g., age, gender, how they would describe their visual impairment) and asked participants about their prior experience with inline and extended descriptions. We then played back three variations of descriptions created from the same written and recorded descriptions using Rescribe: extended descriptions, inline descriptions, and extended-inline descriptions (Figure 4). We randomly selected the order in which we played back the descriptions and asked users to answer three Likert scale questions after each description (Figure 7). At the end, we asked users to compare the three descriptions with respect to preference and when they would use each. We audio recorded and transcribed all interviews.

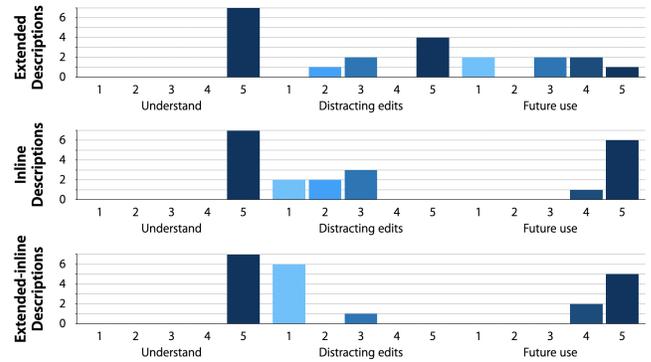

Figure 7. We played participants 3 audio description samples from the same draft description from the same video clip. We randomized the order of the descriptions and after each description asked users to rate their agreement with three Likert scale statements (1-Strongly Disagree to 5-Strongly Agree): I was able to understand the audio descriptions in this clip ("Understand"), I found the editing in this clip to be distracting ("Distracting edits") and I would use this type of audio description in the future ("Future use")

**Participants:** We recruited participants using an email list of blind and visually impaired participants from prior studies that the authors had access to (originally, participants were recruited using social media). Participants ranged from 20-54 years old (3 female and 4 male) and described their visual impairments as blind (6 participants) or low vision (1 participant). All participants used inline audio descriptions (e.g., within movies or TV shows) in the past but only 3 participants had previously listened to extended descriptions.

**Limitations:** Our interview considers first-time participant feedback on a single variation of audio descriptions for a video that they are watching without a particular task in mind (e.g., to note down locations to visit in Ojai). From this study, we cannot know how long term use or how specific task performance might be impacted by description types. Instead, we report initial participant reflections on preference and utility of audio description types generated by Rescribe based on examples.

**Findings**

All participants preferred inline and extended-inline descriptions over extended descriptions when possible (i.e., descrip-tion could be placed with music or ambient sound without overlapping important audio).

**Extended descriptions:** Most participants found extended descriptions to be distracting (Figure 7), with all participants stating that the extended descriptions disrupted the flow of the video. As P3 mentioned, *"it's like technically clear, but sort of, or physically, emotionally jarring"*, and P2 mentioned that *"every time they stopped the music distracted me from the video itself"*. P2 indicated that they did not understand a scene transition until they heard the uninterrupted source audio track. Most participants indicated that while extended descriptions may be acceptable in unavoidable cases (e.g., for necessary educational content, or chaotic scenes), but that *"really, this is a last resort"* (P1).

**Inline descriptions:** Overall, participants indicated that the inline description example, that had been automatically edited down from the longer extended descriptions, met their expectations for audio descriptions such that they would want to use these descriptions in the future (Figure 7). As P5 stated: *"it seemed pretty standard, like what I'm used to"*. The primary reason that they preferred inline to extended descriptions is that they did not disrupt the flow of the music. P1 summarizes: *"[inline descriptions are] much more concise and it's less disruptive of the flow of the, the music track and everything just seems to fit together much better this way"* (P1). While no users rated inline descriptions as distracting (Figure 7), inline descriptions fell between extended descriptions and inline descriptions for user distraction ratings. One of the users who also worked in sound design indicated *"I could hear it chopping words."* (P4), but that the inline description was preferable to pausing the audio as in extended descriptions. P3 also compared all of the techniques favorably to speeding up descriptions: *"I prefer all of these to a thing that they do sometimes of speeding up the descriptions"*. When asked to compare the amount of content between the extended descriptions and the inline descriptions, users noticed small changes they wouldn't mind missing if it could be inline rather than pausing the video (e.g., "mentioning both french fries and tatertots").

**Extended-inline descriptions:** Participants all indicated that they would use extended-inline descriptions in the future, and expressed excitement about the new audio description type as it combined desirable properties of each description type: *"I feel like it's able to combine the best of the extended description with making sure that the video still feels like it flows correctly... I also felt like I could better follow along with the video"* (P6). Most participants stated that they would use extended-inline descriptions when possible, while 2 participants stated that they would like the option between inline and extended-inline descriptions depending on setting (e.g., watching with a sighted person), and content interest. Participants mentioned potential use cases, including when downloading from audio-only media sources (e.g., Blind Mice Vault), and cases when the authors needed to fit in a lot of information: *"It kind of makes me think of like the opening titles to the Simpsons where it has all that stuff happening during the opening theme and, and like the [audio describer] has to talk at like 10 miles a minute to come all in. , I really thought that this one was really good."* (P7). Surprisingly, 6/7 participants strongly disagreed that the editing was distracting (compared to 2/7 for inline descriptions). When following up, participants stated that when listening to the music they either: didn't notice the looping, felt they would not notice if they were not listening for the looping, or did notice but didn't mind the looping.

## EVALUATIONS: CREATING AUDIO DESCRIPTIONS

We created Rescribe to support people in drafting and editing their own audio descriptions. We evaluated Rescribe with users who had video editing experience, but had not created audio descriptions in the past. In our user study, we wanted to find out: ***How do descriptions created with Rescribe differ from descriptions created with traditional editing tools?***

**Materials:** We selected two Creative Commons videos around 10 minutes long from different domains (vlog-style day in the life [12], and travel guide [35]). From each video we, selected two segments for description aiming for approximately an equal amount of gap time in each segment.

**Procedure:** We conducted remote user studies with 8 participants with video editing experience to compare Rescribe with an existing tool. We recruited participants with video editing experience using our institution's Slack channels and e-mail lists. Each participant filled out a pre-study survey about their prior video editing, transcription, and audio description experience. Then, we shared an example of an existing professional audio description, guidelines for writing audio descriptions (see Audio Description Guidelines), and invited users to ask questions. During the task phase, each participant created an audio description for two segments of one video. One segment they audio described using the existing tool, and the other they audio described using Rescribe. The order of the interfaces and the correspondence of interface and clip described were balanced between participants such that each interface ordering, clip ordering combination was used once.

For the existing editing tool, we wanted a workflow that would enable: (a) writing scripts anchored to a timestamp, (b) recording voice-overs for audio descriptions, and (c) timeline-based editing to adjust descriptions after they are recorded. For this purpose, we selected a full-featured video editor (Final Cut Pro) and used markers for scripting and the voice-over feature for recording. Other editors such as Adobe Premiere Pro allow similar functionality. In pilot studies, we tested an audio description specific tool, YouDescribe [25]. However, compared to Final Cut Pro, it did not allow attaching scripts to times on the video track, and adjusting the position and length of recorded clips was difficult compared to other editing software.

For each tool and video clip pair, the participants first received a 5 minute tutorial covering navigation, scripting a description at a video time code, and recording the descriptions. We then asked participants to create their audio descriptions within 12 minutes, by first preparing a script, then recording and refining their descriptions. To make sure participants had time remaining for recording and adjusting, we limited scripting time to 9 minutes. We asked participants to compare the experience of scripting with both tools. Afterwards, participants received their audio description results and compared the two in written comments.

|     | **Final Cut (init.)** | | **Final Cut (final)** | | | **Rescribe (init.)** | | **Rescribe (final)** | | |
|     | # desc | # words | % desc | % words | % overlap | # desc | # words | % desc | % words | % overlap |
| --- | --- | --- | --- | --- | --- | --- | --- | --- | --- | --- |
| P1  | 4  | 28  | 100 | 100 | 50  | 8  | 63  | 100 | 73 | 0 |
| P2  | 14 | 88  | 100 | 100 | 71  | 12 | 90  | 100 | 83 | 0 |
| P3  | 8  | 25  | 100 | 100 | 0   | 8  | 66  | 100 | 74 | 0 |
| P4  | 18 | 73  | 78  | 73  | 5   | 10 | 45  | 90  | 69 | 0 |
| P5  | 5  | 47  | 100 | 100 | 0   | 9  | 103 | 100 | 92 | 0 |
| P6  | 16 | 115 | 100 | 100 | 56  | 12 | 102 | 100 | 91 | 0 |
| P7  | 9  | 48  | 100 | 100 | 78  | 11 | 89  | 64  | 64 | 0 |
| P8  | 26 | 216 | 27  | 26  | 100 | 23 | 211 | 40  | 21 | 0 |
| mean | 13 | 80 | 85  | 87  | 45  | 12 | 96  | 87  | 71 | 0 |
| std  | 7  | 59 | 33  | 25  | 36  | 5  | 47  | 21  | 21 | 0 |

Table 1. We report on the number of descriptions scripted and recorded with Final Cut and Rescribe. Specifically, for both Rescribe and Final Cut Pro user's initial scripted descriptions we report the number of distinct description segments (e.g., either sentence boundaries or manually defined by users), and the number of words.

We audio recorded and transcribed all sessions. We also screen recorded interactions with both interfaces, and instrumented Rescribe to capture interactions.

**Participants:** Participants (P1-P8) ranged from 29-45 years old (3 female, 5 male) and all had prior experience editing videos. While all participants had casual use of some type of video editing tool (e.g., iMovie, Windows Movie Maker) in the past, 4 had experience with professional editing tools (e.g., Premiere, Final Cut Pro), and 2 had experience with transcription software. P1, P3, and P6 had created 1-10 videos in the past, 3 participants (P4, P7, P8) had created 10-50 videos in the past, and P2 and P5 had created over 100 videos in the past.

**Limitations:** As in many studies of new systems, this study is limited to the scope of first-time users of Rescribe, while all participants had experience with timeline-based video editing (and some additional experience with Final Cut Pro). Users also had little to no experience with audio descriptions (listening and creating) and the source material. Familiarity with both tools, audio descriptions as a medium, and the source material (e.g., describing your own media) would likely result in different user behaviour.

### Findings

We report on the benefits and trade-offs of Rescribe and Final Cut Pro with respect to scripting, recording, and final results.

**Scripting and recording:** Users scripted and recorded similar amounts of description content in terms of number of words, with 80 ($\sigma = 59$) words for descriptions scripted with Final Cut Pro and 96 ($\sigma = 47$) words for descriptions scripted with Rescribe. When comparing the two, users mentioned that they valued the fine-grained scrubbing in Final Cut Pro and the navigation to gaps in Rescribe. Users mentioned additional benefits of Rescribe related to the ability to navigate to gaps (P1-P8), the ability to contextualize your audio description with respect to the narration (P5, P8), and the ability to budget time based on how many gaps there are left to describe (P4). Finally, several users mentioned that they liked that the recording option was closely linked to the corresponding script, whereas in Final Cut Pro *"I would lose track of where I was and which marker I was supposed to work on"* (P8). P3 suggested that the tool let you record your descriptions all the way through (their strategy in Final Cut Pro), and edit them after the fact. Most users recorded all of the descriptions that they scripted in both interfaces, but in the cases where they didn't in Final Cut Pro, it was due to time constraints (P8), and editorial decisions to avoid overlapping descriptions (P4).

**Editing:** Most users performed no or minimal editing on both interfaces due to time constraints. When users edited in Final Cut Pro, they primarily deleted or adjusted existing descriptions (P4, P5). In Rescribe, users primarily edited the script (e.g., to adjust wording as they gained more context). In Final Cut Pro, no users edited the script.

**Audio description compositions:** A key problem with Final Cut Pro descriptions is that most participants, 6 of 8, did not successfully fit their descriptions in the gaps provided (Table 1). This led to producing many descriptions completely overlapping one another or the speech in the video (P1, P2, P6, P7, P8), making the speech difficult to understand in the source content or the descriptions (violating the description guideline of avoiding overlapping speech, G4). P4 included only a slight overlap with one speech line. While participants who created significant overlaps generally preferred the descriptions they created without the overlapping content, P2 specified they preferred the content of the Final Cut Pro description and the placement of the Rescribe description. On the other hand, participants who created inline audio descriptions with Final Cut Pro successfully stated that they preferred their Final Cut Pro composition. Participants who successfully manually placed descriptions in Final Cut Pro stated that they would like to see Rescribe improve in placement through more evenly spacing content when space was available *"it felt like the descriptions were squished together"* (P3).

**Audio description as a new task:** Given that all participants were first-time audio describers, many expressed uncertainty through questions about what to describe and how to describe it (e.g., whether to describe text on screen, whether to describe content in an opening montage). During the course of the study, most participants asked whether certain types of content should be described. When reflecting on whether they would publish descriptions, some users confirmed they would be willing to publish their preferred clip (P6, P8, P3) in the current state, or stated things they would change before publishing: *"The description [I created with Rescribe] was richer and better placed"*, but before publishing, they would still want to *"make the description a bit longer and richer."* (P1). Others raised concern with their knowledge of the domain: *" I think I would need to do this 10 times before I felt comfortable making them for real."* (P2).

## EVALUATIONS: INFORMAL EXPERT DEMO

While we originally designed Rescribe with novice audio description creators in mind, we wanted to learn: ***Could Rescribe fit within an expert pipeline for audio descriptions?***

During a 1-hour meeting with 3 audio description professionals (2 of whom we also spoke to before we started the project), we demonstrated Rescribe and shared our results. The professionals all expressed enthusiasm about testing the tool within their workflows and requested that Rescribe output to professional audio editing tools for additional mixing. While professionals indicated Rescribe could be a better replacement for their existing scripting software (currently re-purposed captioning software), the professionals indicated greater excitement about the optimization for placement and shortening than about the visualization of the audio gaps. Two professionals mentioned that while they were initially skeptical about automated techniques (e.g., automatically describing what is important in a scene), they liked Rescribe because it could preserve authoring capability while automating the more tedious iteration step. While professionals remained skeptical about vision-based approaches for scripting assistance, they indicated it would be helpful to identify particular objects that they had already audio described and automatically suggest the relevant descriptions.

## DISCUSSION AND FUTURE WORK

Our work allows users to automatically retarget extended descriptions to inline descriptions improving the ease and efficiency of creating such descriptions without overlapping content. We also find that audio description users expressed interest in using extended-inline descriptions newly enabled by Rescribe. Based on our results, Rescribe could be further improved by integrating with existing expert editing tools and allowing users to explore the parameters. In addition, Rescribe could consider automated feedback to describers to improve initial drafts and confidence in their ability to create high-quality audio descriptions.

**Improving automated methods.** Rescribe offers several areas to improve the efficiency and accuracy of its approach. We manually set parameter weights to balance the terms in our cost function, but learned parameters may produce better results. But, most existing audio descriptions are crafted for movies and TV-shows (as in Rohrbach et al.'s AD corpus [46]) rather than for independently produced content, and extended-inline descriptions do not exist to the best of our knowledge. We hope our tool will enable broader production of video descriptions we may eventually learn from. Also, we rely on text-based speech editing to edit together our candidate audio descriptions. A constraint of our system is that we do not use paraphrasing techniques that require words that do not already exist in the speech (e.g., "An orange car" to "A car"), as synthesizing speech leads to low-quality results when few training examples exist. As synthesized speech improves [30], we will use it. We investigated using automated methods (e.g., dense image and video captioning [26, 29], scene description [51, 33, 21]) to produce descriptions for videos (or keyframes), but current tools provide results that are too general or inaccurate to be useful. In the future, we will investigate expert-appropriate vision approaches, such as recognizing and suggesting descriptions for objects or actions that the describer has previously described (an expert-requested feature).

**Audio description as post-production.** Audio description remains a post-production step rather than an integrated component of the original video production process. Rescribe optimizes audio description placement based on this constraint. Prior work has integrated audio description in the production of live performances [56, 36]. In the future, we will consider how to integrate audio description into video production. For instance, we will suggest video edits similar to the current edit that would make it easier to place audio descriptions. Or, we can recognize visual content in the video that is unexplained in the script (e.g., the video shows sifting flour, but the script says "add flour" without mentioning that it should be sifted). Then, highlight unexplained visual content in the video and corresponding script position, such that a video creator could revise the script before recording it.

**Platform support.** Audio descriptions are currently available on around one third of newly-produced traditional media, but such descriptions rarely exist for platforms with user-generated content (e.g., YouTube, Vimeo). To include AD, users must upload a separate video with a different audio track or use a third party platform such as YouDescribe [25]. Platforms that host video should support, and provide education initiatives, for audio descriptions to increase their availability.

**Audience description preferences.** Existing description guidelines center on traditional media (movies, plays, TV) that feature visually dense settings, action sequences, and interactions. Few guidelines exist for video types that are more common on YouTube (e.g., vlogs, how-tos, reviews). An improved understanding of audience preferences (as newly available for internet images [53, 20] and GIFs [19]), would let us provide better guidance and feedback to new describers.

## CONCLUSION

We present Rescribe, a system that enables people to create narrated, inline audio descriptions quickly for videos without audio descriptions. Our system allows audio description authors to write and record extended descriptions and automatically transform such descriptions to inline descriptions and extended-inline descriptions. We hope that the development of such systems makes videos more widely accessible to all.

## ACKNOWLEDGEMENTS
We thank Kyle Murray for his contributions to the design and development of Rescribe. We also thank the audio description professionals who we talked to for sharing knowledge and design insights that contributed this work.